\documentclass[twocolumn]{svjour3}         
\smartqed  
\usepackage{graphicx}
\usepackage{mathptmx}      
\usepackage[dvips]{epsfig}
\usepackage[dutch,english]{babel}          
\usepackage{fancyhdr}
\usepackage{array}
\usepackage{wrapfig}
\usepackage{subfigure}
\usepackage{graphics}
\usepackage{hhline}
\usepackage{latexsym}
\usepackage{amssymb}
\usepackage{amsmath}
\usepackage{pifont}
\usepackage{palatino}
\usepackage{charter}
\usepackage[charter]{mathdesign}
\usepackage{xr}
\usepackage{caption}

\begin{document}

\title{Modeling final-state interactions with a relativistic multiple-scattering
approximation
}
\subtitle{"Relativistic Description of Two- and Three-Body Systems in Nuclear
Physics",
 ECT$^*$, October 19-13 2009}


\author{W. Cosyn         \and
        J. Ryckebusch
}


\institute{W. Cosyn \at
              Department of Physics and Astronomy,\\
 Ghent University, Proeftuinstraat 86, B-9000 Gent, Belgium \\
              \email{wim.cosyn@ugent.be}             \\
             \emph{Present address:}  Florida International University, Miami,
Florida 33199, USA
           \and
           J. Ryckebusch \at
               Department of Physics and Astronomy,\\
 Ghent University, Proeftuinstraat 86, B-9000 Gent, Belgium \\             
}

\date{Received: date / Accepted: date}

\maketitle

\begin{abstract}
We address the issue of nuclear attenuation in nucleon and pion
knockout reactions. A selection of results from a model based on
a relativistic multiple-scattering approximation is presented.  We
show transparency calculations for pion electroproduction on several
nuclei, where data are in very good agreement with calculations
including color transparency. Secondly, we discuss the density
dependence of reactions involving one or double proton knockout. The
latter reaction succeeds in probing the high density regions in
the deep interior of the nucleus.

\end{abstract}

\section{Introduction}
\label{sec:intro}
It is an open issue how and at what energy scale hadrons emerge
from quarks and gluons, the fundamental constituents of quantum
chromodynamics. The behavior of nucleons in a
nucleus is another topic of great interest and this can be studied by
exploring the limits of the shell-model description of nuclei.  This
 can be achieved, for instance, by looking at the medium modifications of
hadron properties, the density dependence of the nucleon-nucleon
interactions, or the properties of the nucleon-nucleon
correlations at very short internucleon distances.

These fascinating issues can be experimentally explored at
facilities that can probe the nucleonic and sub-nucleonic length
scales, such as Jefferson lab, J-PARC, and FAIR.  Various
aspects of nuclear structure can be studied in reactions such as
$A(e,e'p)$, $A(e,e'\pi)$, $A(p,2p)$, $A(\gamma,2p)$, etc., whereby
a fast leptonic or hadronic probe ejects one or more hadrons from
the target nucleus.  To interpret the data from these experiments, one
needs models that can quantify the effect of nuclear attenuation on
the impinging and ejected nucleons and pions.  These model
calculations can for example be used to identify QCD-mediated
deviations from traditional nuclear physics predictions and to map the
density regions of nuclei contributing to the cross section of
 some specific reaction.

Here, we present a selection of results obtained in a model based on a
relativistic extension of Glauber multiple-scattering theory for the
description of nuclear attenuation on the impinging hadrons
(initial-state interactions, ISI) or ejected hadrons (final-state
interactions, FSI).  This relativistic multiple-scatttering Glauber
approximation (RMSGA) model provides a comprehensive theoretical
framework that can be used for  a variety of leptonic and hadronic
probes and one or more outgoing nucleons and/or pions. The RMSGA
model has no free parameters.  It is used to calculate several
observables like cross sections and nuclear transparencies.  The
general features of the model are introduced in Sec. \ref{sec:model}
and some results are shown in Sec. \ref{sec:results}.

\section{Relativistic multiple-scattering Glauber approximation}
\label{sec:model}

Glauber multiple-scattering theory can be applied when the
wavelength of the scattering particle is small in comparison
with the typical interaction range between the scatterer and
the spectator particles.  Examples of such kinematic conditions
will be discussed in Sec. \ref{sec:results}, with nucleon and pion
energies in the range of a few GeV.  Originating from optics, Glauber
theory uses the eikonal approximation for high energy particles
scattering under small angles.  A relativistic extension of this
eikonal approximations was developed and has been applied to
$A(e,e'p)$ reactions \cite{Ryckebusch:2003fc,Lava:2004zi},
$A(p,2p)$ processes
\cite{VanOvermeire:2006dr,VanOvermeire:2006tk}, pion photo- and
electroproduction reactions \cite{Cosyn:2006vm,Cosyn:2007er} and
neutrino-induced nucleon knockout \cite{Martinez:2005xe}.  Relativity
is accommodated both in the kinematics and dynamics, and all hadrons
involved in the reaction process are described by relativistic wave
functions.  The
impulse approximation is used for the interaction of the beam particle
with the target.

 As an example we mention the Glauber phase in the RMSGA model
entering the amplitude for a reaction with an impinging proton 
which has $i$ hadrons in the final state

\begin{multline} \label{eq:G}
 \mathcal{G}(\vec{b},z)=
 \prod _{\alpha_{occ} \ne \alpha} \int d \vec{r}'
\left| \phi _ {\alpha_{occ}} \left( \vec{r}'     \right) \right|^2\\
\times \left[
1 -  
  \theta \left( z - z' \right) \Gamma_{pN} \left(
\vec{b}' -
\vec{b} \right) \right]\\
\times \prod_i
\left[
1 -  
  \theta \left( z'_i - z_i \right) \Gamma_{iN} \left(
\vec{b}'_i -
\vec{b}_i \right) \right]\,.
\end{multline}
Here, $\alpha$ stands for the quantum numbers of the nucleon that
interacts with the initial probe, $\vec{r}=(\vec{b},z )$ is the
coordinate of the hard interaction, with  $z$  along the momentum
of the impinging hadron. The coordinate
$\vec{r}_{i}=(\vec{b}_ {i},z _{i} )$ is defined such that the $z_{i}$
axis is along the direction of the momentum of the ejected hadron $i$.  The
$\vec{r}'_{i}=(\vec{b} '_ {i},z ' _{i} )$  denotes the coordinate 
$ \vec{r}' $ in the reference system defined by $ \vec{r}_ {i} $.

The product $\alpha_{occ}$ runs over all
residual nucleons, with $\phi _ {\alpha_{occ}} \left( \vec{r}'
\right)$ the single particle bound-state wave functions obtained from
the Serot-Walecka model \cite{Furnstahl:1996wv}. The Heaviside
function ensures that only nucleons in the backward path of the
impinging hadron and nucleons in the forward path of the ejected
hadrons can make contributions to their phase.  Information on the ISI
and FSI is contained in the profile function, and the product $i$ runs
over all particles subject to nuclear attenuations.  The profile
function $\Gamma_{iN}$ for nucleon-nucleon and pion-nucleon scattering
depends on three energy-dependent parameters and has a Gaussian form:
\begin{equation}
\label{eq:gamma}
\Gamma_{iN} (\vec{b}) =
\frac{\sigma^{\text{tot}}_{iN}(1-i\epsilon_{iN})}
{4\pi\beta_{iN}^2}\exp{\left(-\frac{\vec{b}^2}{2\beta_{iN}^2}\right)}\,\;
(\textrm{with} \; i = \pi \; \textrm{or} \; N')\,.
\end{equation}
The three parameters $\sigma^{\text{tot}}_{iN}, \epsilon_{iN},
\beta_{iN}, $ are fitted to nucleon-nucleon and pion-nucleon scattering
data \cite{PDBook,Lasinski:1972tr,Arndt:2003if}.

The scalar Glauber phase $ \mathcal{G}(\vec{b},z) $ is a complex
quantity which encodes the combined effect of attenuation for a
specific 
exclusive reaction that is initiated at the position
$\vec{r}(\vec{b},z)$. The deviation of the norm $ \mid
\mathcal{G}(\vec{b},z) \mid $ from 1 is a measure of the magnitude of
nuclear attenuation.

\begin{figure}
  \includegraphics[width=0.45\textwidth]{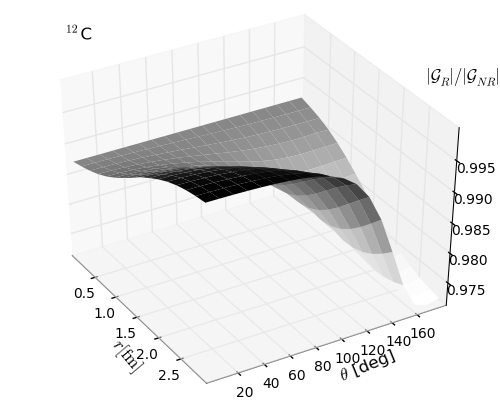}
\caption{The effect of the lower components in the wave functions for
  the scattering centers on the computed Glauber phase $\mathcal{G}$
  at $p$ = 1.5 GeV for proton emission from  $^{12}$C.  The figure
  shows the ratio of the norm of the Glauber phase as computed 
  with the full relativistic single-particle wave function to
  the one which only retains the upper Dirac components. The $
  \mathcal{G}(\vec{b},z) $ is computed for a proton escaping along the
$z$ axis. The coordinates $(r,\theta)$ indicate the position of the
  initial absorption process which triggers the reaction.}
\label{fig:relativity}       
\end{figure}

In our relativistic formulation of Glauber multiple-scattering
theory the single-particle densities $ \left| \phi _ {\alpha_{occ}}
\left( \vec{r}'     \right) \right|^2 $ in Eq.~(1) contain an upper
and lower component. It is conjectured that the effect of relativity
on the magnitude of the nuclear attenuation can be estimated by the
relative contribution from the lower components. 
Fig.~\ref{fig:relativity} shows the effect of the lower components of the
single particle densities on the Glauber phase of an outgoing particle
with an energy representative for the few GeV range.  This shows that
the effect of relativity on the computed magnitude of nuclear
attenuation is rather small, typically a few percent.

It can be expected that the nucleon-nucleon and pion-nucleon profile
functions entering Eq.~(\ref{eq:G}) will be subject to medium modifications
\cite{Bertulani:2010kk}.  Mechanisms like Pauli blocking often lead to
an effective reduction of the nucleon-nucleon cross sections in the
medium. At higher energies the effect of Pauli blocking is
small. Another important source of medium effects are short-range
correlations. Short-range correlations (SRC) can be included in the
treatment of the Glauber eikonal phase by using the information that a
nucleon is present at the coordinate of the hard interaction. 

This can be achieved in the following way.  First, the squared
single-particle wave functions in Eq. (\ref{eq:G}) can be
approximated  by the one-body density of the target nucleus 
$\rho_A ^{[1]}(\vec{r})$ defined as

\begin{multline}
\mid \phi _ {\alpha} ( \vec{r} ) \mid ^{2}  \rightarrow \frac{\rho_A^{[1]}(\vec{r})}{A}=\int
d\vec{r}_2 \ldots \int d\vec{r}_A
\left(
\Psi_A^{\text{g.s.}}(\vec{r},\vec{r}_2,\ldots,\vec{r}
_A)\right)^\dagger\nonumber\\
\times\Psi_A^{\text{g.s.}}(\vec{r},\vec{r}_2,\ldots,\vec{r}_A)\,.
\end{multline}
Here, $\Psi_A^{\text{g.s.}}$ is the ground-state wave function of the target
nucleus, obtained by antisymmetrizing the product of the
single-particle wave
functions $\phi _ {\alpha}$.
This substitution has minor effects on the results obtained in the RMSGA
model \cite{Ryckebusch:2003fc}.  In a next step , the averaged density $\rho_A
^{[1]}(\vec{r})$ can be substituted with the ratio of the two-body density
$\rho^{[2]}_A$ (normalized as $ \int d\vec{r}_1 \int d \vec{r}_2
\rho^{[2]}_A(\vec{r}_1,\vec{r}_2)=A(A-1)$) and the one-body density:
\begin{equation}\label{eq:subs}
  \rho^{[1]}_A(\vec{r}_2) \rightarrow
\frac{A}{A-1}\frac{\rho^{[2]}_A(\vec{r}_2,\vec{r})}{\rho^{[1]}_A(\vec{r})}\,,
\end{equation}
where $\vec{r}$ is the coordinate of the hard interaction.  In the case of an
uncorrelated two-body density:
\begin{equation}
\rho^{[2]}_{A,\text{uncorr.}}(\vec{r}_1,\vec{r}_2)\equiv\frac{A-1}{A}\rho^{[1]}
_A(\vec { r } _1)\rho^{[1]}_A(\vec{r}_2)\,,
\end{equation}
 and Eq. (\ref{eq:subs}) becomes
trivial.  One can include SRC in the two-body density by adopting the following
functional form
 \cite{Frankel:1992er}:
\begin{equation}
 \rho^{[2]}_{A,\text{corr.}}(\vec{r}_1,\vec{r}_2)
\equiv\frac{A-1}{A}\gamma(\vec{r}_1)\rho^{[1]}_A(\vec{r}_1)\rho^{[1]}_A(\vec{r}
_2)\gamma(\vec{r}_2) g(r_{12})\,,
\end{equation}
where $g(r_{12})$ is the so-called Jastrow correlation function and
$\gamma(\vec{r})$ a function that imposes the normalization of the two-body
density obtained as the solution of an integral equation.  With
the above expression for the two-body density , Eq.
(\ref{eq:subs}) becomes
\begin{equation}
 \rho^{[1]}_A(\vec{r}_2) \rightarrow \gamma(\vec{r}_2) \rho^{[1]}_A(\vec{r}_2)
\gamma(\vec{r})
g(|\vec{r}_2-\vec{r}|) \equiv \rho^{\text{eff}}_A(\vec{r}_2, \vec{r})
\, .
\end{equation}
 With above derivation it is clear that the calculation of the FSI
effects can be corrected for SRC by  replacing $\left| \phi _
{\alpha_{occ}} \left( \vec{r}'     \right) \right|^2$ with
$\rho^{\text{eff}}_A(\vec{r} ', \vec{r})/A$ in Eq. (\ref{eq:G})

Colour transparency is a QCD mediated phenomenon that predicts
reduced hadron-nucleon interactions that become more pronounced as the initial
hadron production process occurs at smaller and smaller scales. The
effect of colour transparency can be included in the RMSGA
calculations by making use of the quantum diffusion model of Ref.
\cite{Frankfurt:1988nt} and replacing the total cross section
parameter in Eq. (\ref{eq:gamma}) with an effective one that evolves
from a reduced to its normal value along a certain formation
length $l_h$:
\begin{multline}
\frac { \sigma^{\text{eff}}_{iN}(\mathcal{Z}) }  
{ \sigma^{\text{tot}}_{iN} } =  \biggl\{ \biggl[
 \frac{\mathcal{Z}}{l_h} +
 \frac{<n^2 k_t^2>}{\mathcal{H}} \left( 1-\left(\frac{\mathcal{Z}}{l_h}\right)
\right) \biggr]\\ \times
\theta(l_h-\mathcal{Z}) + 
\theta(\mathcal{Z}-l_h) \biggr\} \,  \; .
\label{eq:diffusion}
\end{multline}
Here, $n$ is the
number of elementary fields (2 for the pion, 3 for the
nucleon), $k_t = 0.350~\text{GeV/}c$ is the average transverse
momentum of a quark inside a hadron, and $\mathcal{H}$ is the hard-scale
parameter (or virtuality) that governs
the CT effect.  For the formation length $l_h \approx 2p/\Delta
M^2$, we adopt the values
$\Delta M^2
= 1~ \text{GeV}^2$ for the proton and $\Delta M^2 = 0.7~\text{GeV}^2$
for the pion.

\section{Results}
\label{sec:results}
\paragraph{Transparencies}
\begin{figure*}
  \includegraphics[width=0.95\textwidth]{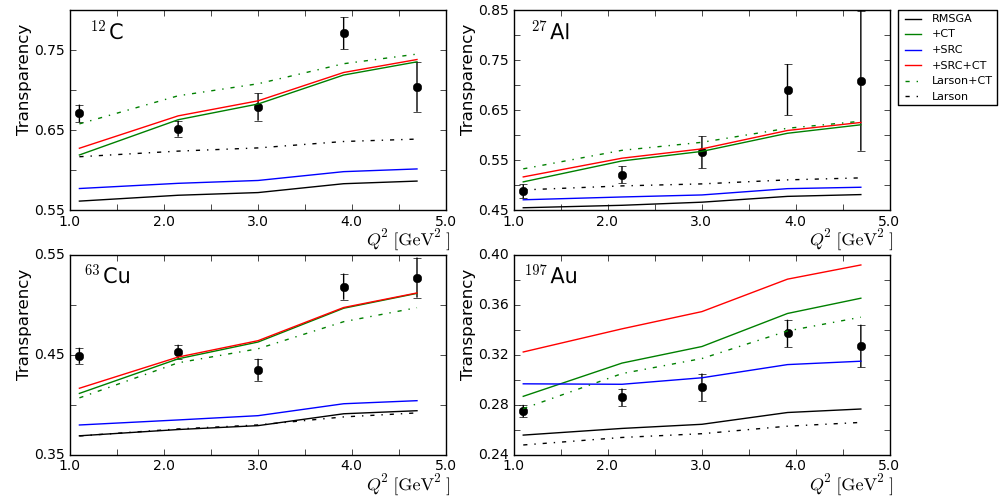}
\caption{\textit{[Colour online]} The $Q^2$ dependence of the nuclear
  transparency for the $A(e,e' \pi^+)$ process in $^{12}\text{C}$,
  $^{27}\text{Al}$, $^{63}\text{Cu}$ and $^{197}\text{Au}$.  The
  dot-dashed curves are the results of the semi-classical model by
  Larson, Miller and Strikman \cite{Larson:2006ge}. JLab data are
  taken from Ref.  \cite{Clasie:2007gq}.}
\label{fig:pionT}       
\end{figure*}

First we show some results for transparency calculations in the RMSGA
model.  The transparency observable is used in the search for the
emergence of partonic degrees of freedom in nuclear reactions and is
defined as the ratio of the cross section on a nucleus to $A$ times
the cross section on a free proton.  Figure~\ref{fig:pionT} presents
the results from our transparency calculations for the pion
electroproduction ($A(e,e'\pi^+)$) reaction on four different nuclei,
together with the experimental data \cite{Clasie:2007gq} and results
from the semiclassical model of Ref. \cite{Larson:2006ge}. The RMSGA
and semiclassical transparencies display a modest increase over the
$Q^2$ range. This behaviour finds a simple explanation in the
$p_{\pi}$ dependence of the $\sigma _{\pi^{+} p}^{\text{tot}}$. The
results contained in Fig.~\ref{fig:pionT} cover a range in pion
momenta given by $2.8 \le p _{\pi} \le 4.4$~GeV. In this range,
$\sigma _{\pi^{+} p}^{\text{tot}}$ displays a soft decrease, which
reflects itself in a soft increase of the nuclear transparency.
After including the effect of SRC (calculations referred to as
RMSGA+SRC), the transparencies are about $2\%$ larger for $^{12}$C, $^{27}$Al
and $^{63}$Cu, and about $4\%$ larger for $^{197}$Au. This reflects
the fact that the medium effectively reduces the free pion-nucleon
cross sections. It is obvious that the effect of SRC does not depend
on the hard-scale parameter $Q^{2}$.  The CT mechanism, on the
other hand, shows a strong $Q^2$ dependence with CT-related
enhancements up to $20\%$ at the highest energies.  These calculations
including CT are in very good agreement with the experimental data.
The results overestimate the Au data somewhat, but the slope is in
agreement with the data.  Upon comparing the RMSGA+CT results
to the semiclassical calculations, we see that the slopes of both
calculations are in excellent agreement, reflecting the use of the
same quantum diffusion parametrization for the CT effect.  There 
are some differences in the predicted values of the transparencies between
the semiclassical and the RMSGA model. The RMSGA predictions are 
somewhat larger for the $^{12}\text{C}$ target, and evolve to
smaller for the $^{197}\text{Au}$ target.  A more recent model
developed by Kaskulov et al.  \cite{Kaskulov:2008ej} also finds
excellent agreement between the calculations including CT and the
data.

\paragraph{Density dependence}

We can also apply the RMSGA model to map which density regions of the
target nucleus are effectively probed in a certain
nucleon-knockout reaction \cite{Cosyn:2009bi}.  The scattering
parameters entering in the Glauber profile function of Eq.
(\ref{eq:gamma}) show little energy dependence for nucleon or pion
momenta above 1 GeV.  This is reflected, for instance, in the soft
energy dependence of the regular RMSGA transparencies in
Fig. \ref{fig:pionT}.  The stronger $Q^{2}$ dependence observed in the data can
be explained by the CT phenomenon. Together with the ability to
deploy the RMSGA framework in a variety of reactions, this allows us
to make statements about the role of nuclear attenuations on the
effectively probed densities for a broad energy range.

For single-nucleon knockout reactions, we can study the density dependence of
the reaction in a factorized approach by calculating $\delta(r,\theta)$, which
represents the contribution to the distorted momentum distribution around $r$
and $\theta$ and is defined as follows:
\begin{eqnarray}
 \rho_{n\kappa } ^{D}(\vec{p}_{m}) & = &
\sum_{s,m} \left|\int d\vec{r}
\frac{e^{-i\vec{p}_{m}\cdot\vec{r}}} 
{(2\pi)^{3}} 
\bar{u}(\vec{p}_m, s) 
\mathcal{G}^\dagger(\vec{r})
\phi_{n\kappa m }(\vec{r})\right|^2 \; , 
\nonumber \\
& = & \iint dr d\theta 
\frac{1}{2} \left[ \sum _{s,m} \left( \left( { D(r, \theta) } \right) ^{\dagger}
\iint dr' d\theta' D(r',\theta')\right. \right. \nonumber\\ 
 &&\left. \left. +{ D(r, \theta) } \iint dr' d\theta' \left( D(r',\theta')
\right)
 ^{\dagger} \right) \right] \; ,   
\nonumber \\
& \equiv & \iint d r d \theta \delta \left( r, \theta \right) \; 
\; .
\label{eq:rhormsga}
\end{eqnarray}
In this equation, $\phi_ {n\kappa m }(\vec{r})$ is the single-particle wave
function of the struck nucleon (with quantum numbers $(n\kappa m)$), the
missing
momentum $\vec{p}_m=\vec{q}-\vec{p}$  is defined as the difference between the
momentum transfer and the final nucleon momentum, and the function
$D(r,\theta)$ is defined as
\begin{equation}
 D(r, \theta) = \int d\phi r^2 \sin{\theta} \frac{ 
 e^{-i\vec{p}_{m}\cdot\vec{r}}
} 
{(2\pi)^{3}} 
\bar{u}(\vec{p}_m, s) 
\mathcal{G}^\dagger(\vec{r})
\phi_{ n \kappa m } (\vec{r}) \; .
\end{equation}
In the factorized approach, the cross section is proportional to the distorted
momentum distribution $\rho_{n\kappa } ^{D}(\vec{p}_{m})$, and
$\delta(r,\theta)$ thus provides a measure for the contributions to the cross
section of the different density regions in the nucleus.

\begin{figure*}
  \includegraphics[width=0.95\textwidth]{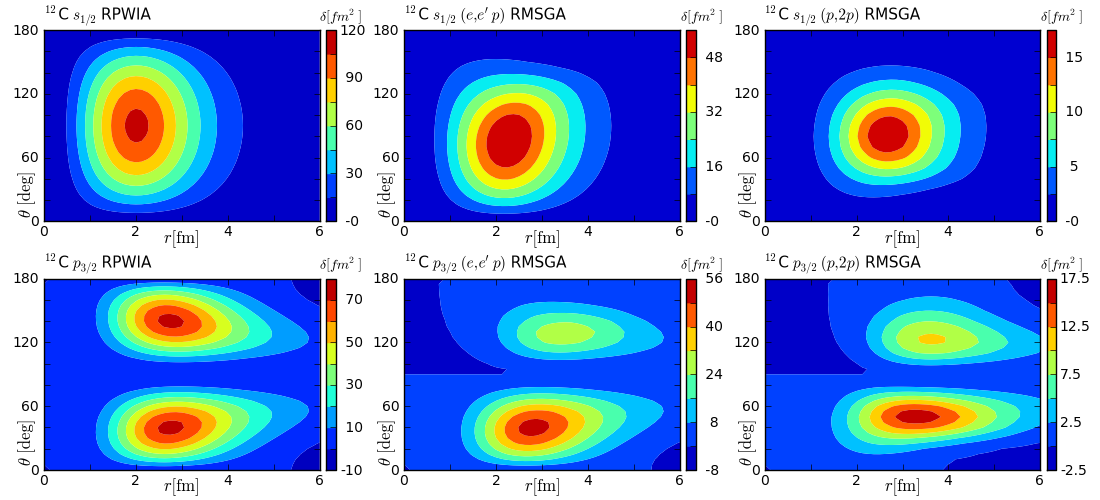}
\caption{\textit{[Colour online]} The function $\delta (r, \theta) $
  for the $^{12}$C$(e,e'p)$ and $^{12}$C$(p,2p)$ reactions. The energy transfer
for both reactions is 1.5 GeV and the
  three-momentum transfer $\vec{q}$ is tuned to probe the maximum
  of the momentum distribution (i.e. $p_m$=0 MeV for knockout from the
  $s1/2$-orbit and $p_m$=115 MeV for removal from the $p3/2$-orbit).
  The proton is ejected along $\vec{q}$ for the $(e,e'p)$ results.  For the
$(p,2p)$ calculations, both protons in the final state have 1.5 GeV kinetic
energy and are detected under an angle of $32^\circ$ on opposite sides of the
incoming proton.  This particular kinematic situation is often
referred to as coplanar and symmetric kinematics.  For the
  sake of reference, the proton root-mean-square radius in $^{12}$C as
  determined from elastic electron scattering is $\left< r^2 \right>
    ^{1/2} = 2.464 \pm 0.012 $~fm \cite{Reuter:1982zz}. }
\label{fig:1nucl}       
\end{figure*}

Results for $\delta(r,\theta)$ for the $^{12}\text{C}(e,e'p)$ and
$^{12}\text{C}(p,2p)$ reactions are shown in Fig.~\ref{fig:1nucl} with
$\theta$ measured from the direction of momentum transfer and $r$
denoting the distance relative to the center of the target nucleus.
The energy transfer is 1.5 GeV and the polar angle of the ejected
protons are determined such that the  kinematics probes the
maximum of the undisturbed momentum distribution. In the relativistic
plain-wave (RPWIA) limit, no ISI or FSI are present
($\mathcal{G}(\vec{r}) \equiv 1)$ and $\delta(r,\theta)$ is equal for
$(e,e'p)$ and $(p,2p)$.  For the selected kinematics described in
the caption to Fig.~3, there is a symmetry-axis for
$\theta=90^\circ$. In the absence of any effect stemming from
nuclear attenuation the forward and backward hemispheres equally
contribute. After including the FSI and/or ISI in the RMSGA
approach, we observe that the nuclear attenuations reduce the value
of $\delta(r,\theta)$, shift the maximum values to higher values of
$r$, and also induce an asymmetry in the $\theta$ direction.  The
biggest contributions to the cross section stem from the forward
hemisphere. The relative contribution from the interior
regions of large target-nucleus density 
gets reduced due to its sensitivity to strong FSI
mechanisms. All these effects are more pronounced in the $(p,2p)$
reaction, where three (one incoming, two outgoing) protons are subject
to the attenuations and the biggest contributions are close to the
nuclear surface.

To formulate the equivalent of Eq. (\ref{eq:rhormsga}) for two-nucleon knockout
reactions such as $A(\gamma,pp)$ in a factorized approach, we have to assume
that the proton pair resides in a relative $S$-state.   This is a reasonable
approximation as
investigations of the $^{16}\text{O}(e,e'pp)$ reaction at the electron
accelerators in Mainz \cite{Ryckebusch:2003tu,Rosner200099} and Amsterdam
\cite{Onderwater:1997zz,Onderwater:1998zz,Starink:2000wc} have clearly shown
that pairs of protons are solely
subject to
short-range correlations when they reside in a relative $S$ state under
conditions corresponding with relatively small c.m. momenta $P$ (or, the initial
protons are very close and moving back-to-back).  This assumption allows us to
write a distorted momentum distribution for the nucleon-nucleon pair:
\begin{multline}
\rho _{n_1 {\kappa}_1, n_2 {\kappa}_2} ^{D} (\vec{P}) =
\sum_{s_1, s_2, m_1, m_2}
\left|  \int d \vec{R}
\frac { 
 e^{-i\vec{P}_m \cdot \vec{R}}
} 
{(2\pi)^{3}} \bar{u}(\vec{p} + \frac {\vec{P}} {2}, s_1)\right.\\
\left. \times \phi_{n _{1} \kappa _1 m_1} (\vec{R}) 
\bar{u}(-\vec{p} + \frac {\vec{P}} {2}, s_2) 
\phi_{n_{2} \kappa_2 m_2} (\vec{R})
\mathcal{G}^\dagger(\vec{R})
\right|^2 \\
 \equiv \iint dRd\theta \delta \left( R, \theta \right) \; 
\; ,
\label{eq:rho2rmsga}
\end{multline}
with $\vec{P}$ ($\vec{p}$) the center of mass momentum (relative
momentum) of the outgoing nucleon-nucleon pair.  The missing momentum $\vec{P}_m
= \vec{P} - \vec{q}$ can be interpreted in the quasi-free approximation as the
center of mass momentum of the correlated pair before the photon interaction.  

\begin{figure}
  \includegraphics[width=0.47\textwidth]{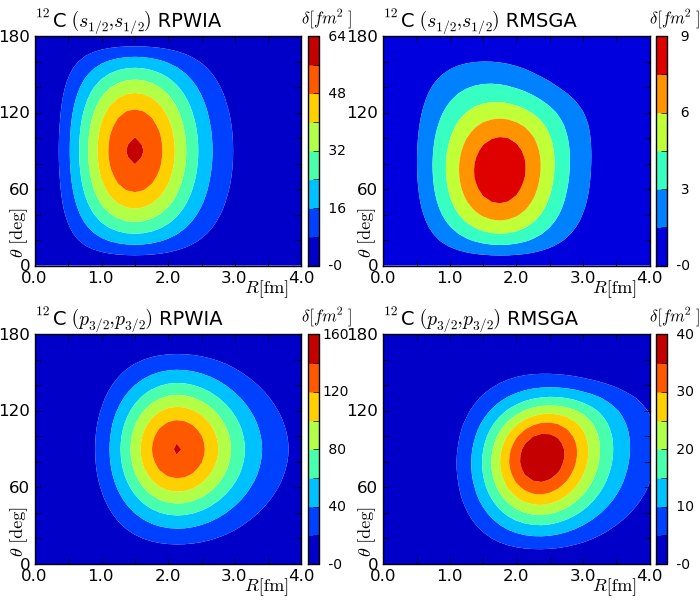}
\caption{\textit{[Colour online]}  The function $\delta (R, \theta) $
  for the exclusive $^{12}$C$(\gamma, pp)$ cross section. We consider an energy
transfer of 3 GeV and a
  three-momentum transfer $\vec{q}$ that is tuned to probe the maximum
  of the momentum distribution $\rho _{n_1 \kappa _1, n_2 \kappa _2} (\vec{P})
  $ (i.e. $P$=0 MeV for knockout from the $(s1/2 -s1/2)$- and $(p3/2 -
  p3/2)$-orbits). We consider coplanar and symmetric kinematics, i.e.
  the two escaping protons have the same energy and polar angle
  $\theta_{pq}$, but escape from the opposite side of $\vec{q}$}
\label{fig:2nucl}       
\end{figure}

Fig. \ref{fig:2nucl} shows $\delta(R,\theta)$ for the dual proton
knockout on $^{12}$C, with proton kinetic energies of 1.5 GeV
in the final state.   We can again quantity the effect of FSI by comparing the RMSGA
to the RPWIA results. It is clear that the biggest values of
$\delta(R,\theta)$ are situated much closer to the center than for the
one-nucleon knockout reactions.  Consequently, the $A(\gamma,pp)$
reaction succeeds in probing the high density regions of the target
nucleus.

\section{Conclusion}
We have shown a selection of results obtained in a model based on relativistic
multiple-scattering Glauber scattering theory.  The model has no free
parameters and can be applied to variety of reactions, with
leptonic and hadronic beams and outgoing nucleons and/or pions.  
Our calculations show that relativity plays a rather modest role in
the magnitude of nuclear attenuation.
We showed very
good agreement of pion transparencies with calculations including colour
transparency.  Secondly, we exploited the robustness of the model to 
 explore which target-nucleus densities can be effectively probed in
knockout reactions involving one, two and three protons. We find
that  the
$A(\gamma,pp)$ reaction probes the interior of the target nucleus, the
$A(p,2p)$ is rather peripheral, whereas the $A(e,e'p)$ 
is somewhat intermediate between these two.

\begin{acknowledgements}
This work was supported by the Research Foundation Flanders and the
Research Board of Ghent University.
\end{acknowledgements}

\bibliographystyle{spphys}

\end{document}